# Emotion Diffusion in Real and Simulated Social Graphs: Structural Limits of LLM-Based Social Simulation

**Qiang Qiqi**

## 1.Introduction

Social media has become an important platform for information dissemination, opinion shaping and emotion diffusion. When public events occur, the spread of emotions on social media may affect social stability, policy decisions, and public perceptions. For example, political events may lead to extreme emotional antagonism, disaster events may trigger panic, and entertainment events may promote emotional contagion. The study of emotion propagation mechanisms in social media is not only of great theoretical significance for Computational Social Science, but also of great value for practical applications such as information dissemination, public opinion management, and crisis response.

In recent years, the application of Large Language Models (LLM) in social media simulation research has gradually attracted attention. Researchers have begun to use LLM to generate virtual social media interaction data to explore the mechanisms of public opinion evolution and emotion propagation in social media. However, it is unclear whether LLM-generated data can realistically simulate emotion diffusion patterns on social media, and how different types of public events affect emotion diffusion remains an open question.

A community is usually defined as a group of people who share common interests, goals, or social ties. Members of a community form close ties through social interaction, information sharing, and emotional resonance. In this environment, emotion communication is special and important. Community emotional communication not only affects individual emotional states, but may also have a profound impact on the overall community climate, opinion formation and social behavior. Therefore, it is important to study the mechanisms, models, and analysis methods of emotion propagation within a community to understand social networks, online opinion, and social psychology.

The purpose of this review is to systematically collate the current state of research in this area and to propose directions for future research. Specifically, this study aims to (1) collate the major research advances in social media community emotion diffusion, and explore its theoretical basis and analytical methods; (2) assess the potential and challenges of LLM in modelling social media community emotion diffusion, and analyze its strengths and shortcomings in terms of the quality of the generated data, and the matching degree of diffusion patterns; and (3) compare the characteristics of emotion diffusion of different types of public events on different communities, to Reveal the key factors affecting the emotion diffusion mode. Through the analyses in this review, we hope to provide a clear theoretical framework for subsequent studies and provide empirical evidence for the simulation of social emotion diffusion using LLM.

## 2.Literature Review

## 2.Literature Review

### 2.1 Emotional Communication Theory in Social Media Communities

Social media plays a significant role in emotional communication where people tend to express themselves. People exposed to different emotional occasions will have distinguished emotional expressions. Through the emotional contagion theory, people's emotion will be affected by others. This theory is also applicable to social media communities. Emotional contagion is understood as an automatic and unconscious process where individuals regulate their emotions by observing and imitating others' emotional expressions (e.g., facial expressions, tone of voice, body language) (Hatfield et al., 1994). In addition, the Opinion Leader Effect (OLE) is also an important factor influencing the spread of community emotions. Opinion leaders occupy a central position in the community and can effectively influence the emotions and attitudes of other members (Katz & Lazarsfeld, 1955). These theories provide a foundation for research on community emotional communication and help explain differences in emotional communication across different community structures.

Scholars have explored the applicability of these theory beyond traditional settings, particularly in the context of social media. It is found that emotions on social media can be transmitted without face-to-face interaction, and that the expression of emotions affects the user's own emotional state (Ferrara and Yang, 2015). As the most frequently used application, Facebook attracted many people post their emotional state on it, and researches on this are also popular. A large-scale Facebook experiment demonstrated that reducing either positive or negative content in users' news feeds influenced the emotional tone of their subsequent posts (Kramer, Guillory and Hancock, 2014). The study explored how Instagram browsing affects subjective well-being and found that when users viewed more positive posts, their positive emotions increased and their negative emotions decreased, supporting the theory of emotional contagion. This result was independent of the effect of social comparison (Choi and Kim, 2021).

There are different dissemination paths for different emotions. Research shows that negative emotions spread faster than positive emotions, but positive emotions are more acceptable to users(Ferrara and Yang, 2015). People in difficult period will also be influenced by negative posts on social media. During the COVID-19 pandemic, the widespread diffusion of negative emotions on social media had a significant impact on public mental health (Lu and Hong, 2022). Some researchers also found that the influence of negative posts even led to obsessive-compulsive disorder (Wheaton, Prikhidko and Messner, 2021).

### 2.2 Models of information dissemination

The Susceptible-Infected-Recovered (SIR) model, originally developed to describe disease transmission, was later adapted to analyze the diffusion of information and emotions in social media. The SIR model divides social network users into:

- Susceptible (S): users who have not yet been affected by emotions.

- Infected (I): users who have expressed or transmitted a specific emotion.
- Recoverees (R): users who no longer transmit the emotion.

It was found that hub users have a high degree of influence in the SIR propagation process, and their emotional state can spread rapidly in the network. The Independent Cascade Model (IC) assumes that information or emotions spread probabilistically, with each infected individual having a certain probability of influencing its neighboring nodes. In the context of social media sentiment propagation, the model is used to simulate how influencers diffuse sentiment through their fan base.

Building on these models, researchers have integrated the SIR and IC models to simulate emotion dissemination on social media. The Emotional Independent Cascade (eIC) model was introduced to more accurately simulate emotional diffusion. Twitter data was used to analyze the role of different interaction mechanisms (retweets, mentions, comments) in emotional diffusion. Studies have shown that hub users play a crucial role in the SIR propagation process, as their emotional states can rapidly influence a large portion of the network. Research indicates that tightly interconnected social communities are prone to developing homogeneous emotional trends, which can accelerate the spread of both positive and negative emotions (Xiong *et al.*, 2018).

The Linear Threshold Model (LT Model) suggests that an individual's emotional state is controlled by the ‘social influence threshold’ in a social network, i.e., a person will be affected when enough of his or her friends around him or her express a certain emotion. In the social media environment, the herd effect tends to exacerbate the transmission mechanism of the LT model.

Emotion diffusion in social media is a complex process involving multiple diffusion models and social network structures. The SIR, IC, and LT models provide analytical frameworks from different perspectives to help researchers understand how emotions diffuse across social platforms. In addition, small-world networks and opinion dynamics models further reveal the impact of social relationships and group behaviour on emotion diffusion. However, existing studies primarily emphasize short-term emotional diffusion and overlook the effects of information cocoons and extreme emotional polarization.

### 2.3 Social Media Communities Sentiment Analysis

With the rapid development of social media, people express their emotions on social platforms in increasingly rich ways, which provides a massive data source for sentiment analysis. Social Media Sentiment Analysis (Sentiment Analysis) is mainly used to automatically identify and classify emotional tendencies (positive, negative or neutral) in text, images, and videos. The technique is widely used in scenarios such as brand opinion analysis, government decision-making, and crisis response. In recent years, researchers have adopted various natural language processing (NLP) and deep learning techniques to improve the accuracy of sentiment analysis, and have also begun to explore the influencing factors of sentiment propagation and its temporal dynamics.

Three primary methods are commonly employed in social media sentiment analysis. One widely used approach is the lexicon-based method. This method is a sentiment analysis method based on sentiment dictionaries (e.g. VADER, TextBlob). Two widely used lexicon-based tools for social media sentiment analysis are:

- VADER (Valence Aware Dictionary and sentiment Reasoner): designed for social media texts, capable of handling social media features such as emoticons, capital letters, and so on.
- TextBlob: a rule-based NLP library for short text sentiment classification.

While easy to implement, this method struggles with detecting sarcasm, contextual dependencies, and nuanced sentiment expressions, as it primarily relies on word-matching techniques.

The second widely adopted method is the machine learning approach. This method utilizes feature engineering and classification algorithms (e.g. SVM, Random Forest) for sentiment classification. Commonly used methods include Support Vector Machine (SVM) and Random Forest methods. Support Vector Machines (SVM) are particularly effective for small datasets, as they better capture the emotional characteristics of textual content and Random Forest methods improve the stability and robustness of the model by voting through multiple decision trees.

The most advanced and effective approach is deep learning-based sentiment analysis. Deep learning methods are able to automatically extract text features and are particularly suitable for sentiment analysis of short social media texts. BERT (Bidirectional Encoder Representations from Transformers) and RoBERTa perform well in sentiment classification tasks and are able to handle contextual information effectively(Wang *et al.*, 2022). To enhance the accuracy of sentiment analysis on social media texts, researchers have proposed an integrated hybrid deep learning model that combines RoBERTa, LSTM, BiLSTM, and GRU to improve text representation and classification precision. Experimental results indicate that this approach outperforms traditional machine learning methods in social media sentiment analysis and significantly enhances classification accuracy(Tan *et al.*, 2022). Bidirectional Long Short-Term Memory (BiLSTM) is an extension of LSTM (Long Short-Term Memory), which processes input data simultaneously through two LSTM networks, forward and backward, to make fuller use of contextual information, and is especially suitable for tasks that require long-distance dependency, such as natural language processing (NLP),especially for social media text. Aiming at the difficulty of traditional methods in effectively capturing complex semantics in short texts, Long proposes a hybrid model combining a bidirectional long-short-term memory network (BiLSTM) and a multi-head attention mechanism (MHAT) to enhance the performance of sentiment classification in social media texts. The BiLSTM network learns the contextual dependencies of the text through forward and backward propagation to enhance the model's ability of capturing long sequences of dependencies, while the MHAT mechanism computes information in different subspaces through multi-head self-attention to highlight key information expressions in the text. Experimental results show that the method outperforms traditional machine learning methods on social media text datasets and significantly improves classification accuracy, demonstrating the effectiveness of the combined BiLSTM-MHAT strategy in sentiment analysis tasks (Long et al., 2019). This study provides new ideas for the application of deep learning in the field of sentiment computing, especially for the social media text analysis task with variable language expressions and significant implicit features of sentiment.

### 2.4 Large Language Model Stimulation of Social Media Communities

In recent years, Large Language Models (LLMs) have been gradually applied to simulation studies of social media sentiment propagation, demonstrating significant potential. This study demonstrates the potential of combining LLMs with computational agents to simulate human-like behavior, offering a novel architecture and interaction paradigm for building believable interactive applications(Park *et al.*, 2023). Studies have demonstrated its applicability in modeling human behavioral responses. Törnberg et al. (2023) used LLMs in conjunction with Agent-based modelling to create virtual social media environments to study the effects of different news recommendation algorithms on user interaction and dialogue quality, and found that specific algorithms (such as the ‘bridging algorithms") can effectively facilitate constructive discussions across political positions. (Törnberg *et al.*, 2023). The $S^3$ social network simulation system integrates LLM's reasoning capabilities with social-emotional communication modeling, validating its effectiveness in simulating group emotional interactions(Gao *et al.*, 2023). In addition, Ahnert et al. demonstrated the advantages of LLM in capturing long-term sentiment trends in social media by developing Temporal Adapters (TAMs), and found that LLM-generated data could correspond with actual social survey data to a certain extent(Ahnert *et al.*, 2025).

Furthermore, LLMs can enhance the accuracy of time-series analysis. Building on this research, this approach can be applied to social media text analysis to track emotional fluctuations during emergencies (Wang *et al.*, 2024). The important innovation of this study is the first systematic implementation of the deep fusion of structured numerical time series data and unstructured news text, and the proposal of a dual-agent system based on the Large Language Model (LLM): the Reasoning Agent and the Evaluation Agent. The Reasoning Agent is responsible for causal inference and dynamic filtering of news content, while the Evaluation Agent guides the iterative improvement of the reasoning process through feedback of prediction errors. This approach effectively utilizes the inference and reflection capabilities of LLM and overcomes the limitations of traditional prediction models that are difficult to capture complex social event information. In terms of methodological principles and experimental design, this framework possesses clear potential for applicability to the study of social media emotion communication, as demonstrated by the following:

First, LLMs are well-suited for processing and analyzing unstructured textual data (e.g., tweets, comments) on social media. Second, the system has good dynamic adaptability and causal reasoning ability, which can continuously identify and evaluate important events affecting sentiment propagation, and achieve self-optimization of the model through an iterative reflection mechanism. The experiments in And's thesis demonstrate that this deep fusion of textual and time-series data has already significantly improved the prediction accuracy in a number of fields, such as energy demand prediction, exchange rate fluctuation analysis, and bitcoin price prediction, and is therefore theoretically also suitable for sentiment prediction. This model is also theoretically applicable to the field of emotion propagation prediction.

However, despite the significant advantages of LLM in modelling social media sentiment propagation, there are still many limitations. There are significant textual feature differences between LLM-generated data and real social media data. Informal language, extreme emotional expressions and unstructured text features common in real data are often weakened in LLM-generated data. Liu et al.'s study also confirms that although LLM performs well in the rumor detection task, it is still inadequate in handling highly emotional content(Liu *et al.*, 2024). Meanwhile, Diao et al.'s study emphasized that real social media emotion propagation patterns have complex dynamics, while LLM-generated data generally lacks such diverse propagation patterns (Diao et al.,2021).

Moreover, LLM still faces challenges in reflecting the dynamic nature of real social networks. Gao et al. pointed out that LLM simulation systems are usually based on short-term textual reasoning as well, which is unable to effectively capture the long-term user behavioral dynamics in social media emotion propagation. Moreover, variations in information flow algorithms across social media platforms further complicate the realism of LLM-based simulations (Gao *et al.*, 2023).

LLMs often exhibit biases in various contexts. First, the inherent bias of LLM training data affects the authenticity and diversity of the generated content. Peters and Matz pointed out that LLM has gender and age bias in predicting the psychological traits of social media users, and this bias may lead to the distortion of the social media interactions it generates(Peters and Matz, 2024). The trained corpus can lead to biased models of grand prognostications. Wald & Pfahler (2023) investigated how large-scale language models reproduce bias in social media data. By fine-tuning GPT-Neo 1.3B on six social media datasets and measuring sentiment and toxicity values across different populations, they found that the type and intensity of bias varied across datasets(Wald and Pfahler, 2023). Additionally, LLMs tend to exhibit stronger negative biases when processing certain socially sensitive topics. Mei et al. (2023) investigated LLM bias against 93 socially stigmatised groups in a sentiment classification task and found that LLM-generated content tended to be more negatively biased than real social media data, especially when it came to topics such as illness, disability, and educational attainment(Mei, Fereidooni and Caliskan, 2023). To mitigate biases in LLMs, Da et al. (2024) proposed employing causal mediator analysis and counterfactual training to enhance fairness in sentiment analysis(Da *et al.*, 2024).

## 3.Methodology

This study proposes a four-stage framework to systematically compare emotion diffusion processes in real-world and simulated social graphs. The framework includes: (1) a baseline emotion propagation simulator, (2) LLM-driven contagion simulation, (3) real-world diffusion graph construction, and (4) GNN-based emotional state prediction. The modular structure allows for both controlled experimentation and empirical validation.

**Stage 1: Baseline Emotion Propagation Simulator**

We construct a synthetic undirected graph G = (V, E), where nodes V represent users and edges E represent social interactions. Each node v∈V in V is initialized with three attributes: (1) an emotional state $e_v \in \{\text{positive,neutral,negative}\}$, (2) a credibility score $c_v \in [0,1]$, and (3) a susceptibility coefficient $s_v \in [0,1]$. Edges may include types (e.g., comment, mention) and weights.

Three propagation strategies are implemented:

- **Random**: each activated node propagates its emotion to neighbors with a fixed probability;
- **Theory-based**: propagation probability is determined by the sender's credibility, emotional intensity, and receiver's susceptibility;
- **Enhanced IC (eIC)**: incorporates interaction type and node centrality (e.g., hub score) into the cascade function.

The propagation outcome is evaluated using a composite reward function:

- $R_{spread} = \log(1 + |V_{infected}|)$: encourages wide spread;
- $R_{polar} = -\text{Var}(\{e_v\})$: penalizes emotional polarization;
- $R_{cred} = \sum_{v \in V_{infected}} c_v$: rewards credibility-weighted spread.

The simulator records propagation chains, emotional updates, and reward scores at each round for downstream comparison.

**Stage 2: LLM-Driven Emotion Contagion Simulation**

To simulate human-like emotional contagion, we incorporate large language models (LLMs). For each interaction $u \rightarrow v$, a prompt is generated using node and interaction metadata, and an autoregressive LLM (e.g.,deepseek ) produces a natural-language reply on behalf of v.

The generated reply is passed through a sentiment classifier (e.g., VADER or a zero-shot classifier) to infer v's emotional state. Node v is then updated with this label, and the process proceeds recursively.

All LLM-generated chains are logged with text content, emotion labels, and propagation structure, enabling direct comparison with baseline simulations in terms of diffusion depth, emotional drift, and contagion breadth.

**Stage 3: Real-World Social Graph Construction**

Retrieving real-world social graphs from Reddit using the PRAW API and selecting a target subreddit (e.g., r/worldnews), extract posts, comments, user IDs are essential initially. After that the reply structure, a directed graph $G_{real} = (V, E)$ is constructed:

- Nodes V: users who posted or commented;
- Edges E: reply relationships (directional);
- Node attributes: posting frequency, inferred emotion, account age;

- Edge attributes: depth of interaction, text length.

Each comment/post is labeled using a sentiment analysis pipeline to estimate $e_v$, and the temporal reply structure is preserved to reconstruct diffusion chains. This structure enables structural and behavioral comparison with simulated graphs.

**Stage 4: GNN-Based Emotion State Prediction**

Graph Neural Networks (GCN, GraphSAGE) are trained to predict nodes' future emotional states based on current graph snapshots. Input features include:

- Node: initial emotion (one-hot), credibility, post frequency;
- Edge: interaction type, weight.

Models are trained separately on $G_{real}$ and simulated graphs. The prediction task targets final emotional labels. Evaluation metrics include accuracy, F1-score, and macro-F1. Cross-domain generalization is also assessed to evaluate robustness between real and generated graphs.

## 4.Results and Findings

### 4.1 Comparison of Emotion Diffusion Strategies in Simulated Networks

In a randomly generated network structure, we compared three emotion propagation strategies from a single source node. the Random strategy successfully influenced 2 neighboring nodes (nodes 1 & 8) in a network with 26 edges in the graph and an average degree of 5.20, obtaining a total reward of 2.139; the Theory strategy failed to propagate in a sparser graph (only 16 edges, average degree 3.20 ) fails to propagate with a reward of 0.000, while the eIC strategy successfully infects node 4 in a graph with 15 edges and an average degree of 3.00, with a reward of 0.533. This result shows that the density of the graph structure significantly affects the propagation effectiveness, with the Random strategy having a higher penetration in high-connectivity networks, whereas the strategies relying on the probability of propagation modelling (e.g., Theory vs. eIC) are more likely to fail in low-density graphs are more likely to fail.

Further, the overall diffusion performance of each strategy in 50 independent simulations is counted. As shown in Table 1, the Random strategy has an average number of spread nodes of 2.58 and an average reward value of 2.556, which is the best performance among all strategies; the Theory strategy has an average number of spread nodes of 1.44 and an average reward of 1.404; and the eIC strategy has 1.12 and 1.175, respectively. the maximum number of spread nodes of the three strategies are 8, 7, and 5, and the maximum rewards are 8.207, 7.069, and 5.492, and the minimum propagation is 0 for all of them, indicating that propagation failed to occur in some experiments. Overall, the Random strategy is more suitable for modelling the propagation of emotions in complex networks, while Theory and eIC are more sensitive to graph structure and initial conditions.

Table 1:The result of different strategy

| **Strategy** | **Avg. Spread** | **Avg. Reward** | **Max Spread** | **Min Spread** | **Max Reward** | **Min Reward** |
|---|---|---|---|---|---|---|
| **Random** | 2.58 | 2.556 | 8 | 0 | 8.207 | 0.000 |
| **Theory** | 1.44 | 1.404 | 7 | 0 | 7.069 | 0.000 |
| **eIC** | 1.12 | 1.175 | 5 | 0 | 5.492 | 0.000 |

Figure 1 shows visualizations of emotion propagation under three strategies. In the Theory-based model, the seed node failed to spread emotion due to insufficient susceptibility or sparse connectivity. The Random strategy, in contrast, successfully activated two neighbors regardless of node attributes. The eIC strategy resulted in one successful diffusion instance, indicating its potential to model realistic, selective emotion contagion where both node and edge properties matter.

Figure 1: Emotion Diffusion Under different Propagation Strategies

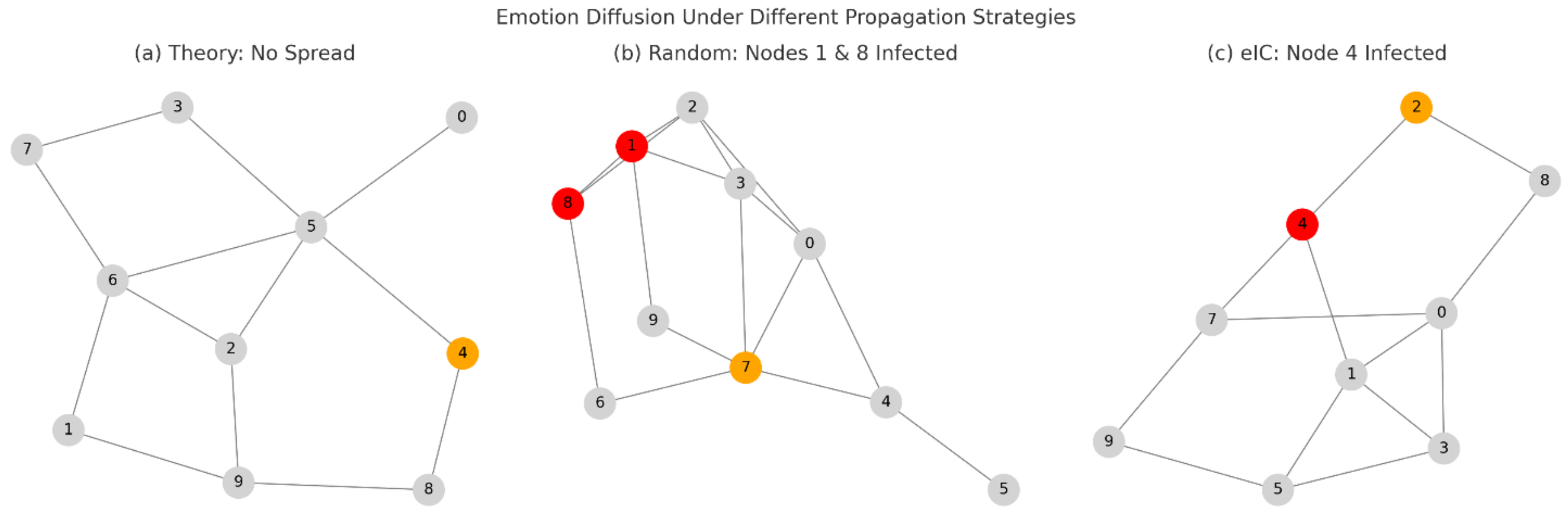

### 4.2 LLM's Emotion Diffusion

To explore the dynamics of LLM-driven emotion diffusion, we initiated a simulation from a single source node (Node 31), which expressed a clearly neutral emotion. The resulting propagation network is visualized in *Figure X*, where node colors represent the round of emotional "infection", and red edges indicate active propagation paths.

In the first diffusion round, Node 31 generated 31 replies, each prompted with instructions to maintain a neutral tone. However, sentiment classification of the generated replies revealed a notable emotion shift:

- 83.9% (26/31) of the replies were classified as *positive*
- Only 16.1% (5/31) were correctly recognized as *neutral*

In the second round, replies continued from newly activated nodes, and the positivity trend became even more pronounced—virtually all responses were labeled as positive. This cascading effect suggests a systematic *positivity bias* in the language model's outputs, even under neutral prompting conditions.

This sentiment drift significantly influenced the diffusion network’s overall emotion structure. The graph rapidly lost diversity in affective tone, forming a homogeneously positive cascade. As a result, downstream tasks such as graph-based sentiment classification struggled with class imbalance: the GCN classifier completely failed to learn meaningful patterns for the *neutral* class, yielding an F1 score of 0.

Figure 2 illustrates the infection routes and node activation rounds, highlighting the central role of the initial node and the strongly radial pattern of spread. This finding underscores a critical limitation of current LLMs in multi-turn, networked conversation settings: they tend to amplify affective polarity, even when such polarity is not present in the input.

Figure 2:Emotion Spread Visualization by Infection Round

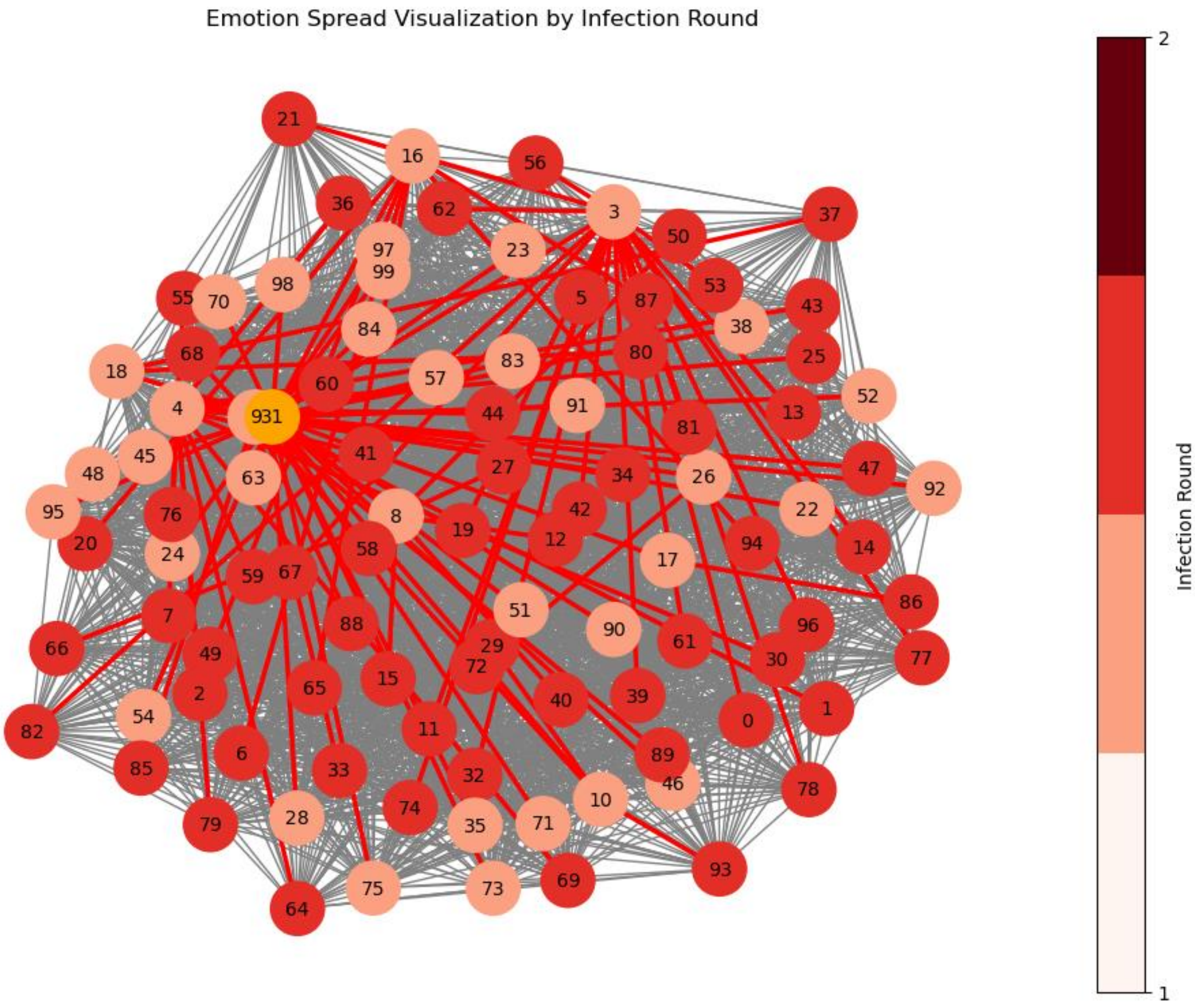

## 4.3 Real world data and emotion diffusion

To facilitate the modeling of emotion diffusion in real-world social networks, we constructed a high-quality interaction dataset derived from the **Reddit r/worldnews** subreddit. This subreddit was chosen due to its active user base, rich comment hierarchies, and diverse range of conversational topics, making it ideal for studying multi-user emotional interactions.

The collected dataset comprises three core components: user-level interaction links (with each record capturing a *source_user* and *target_user* to reflect directed communication), thread identifiers (*post_id* indicating the origin thread of the interaction), and the raw textual content of user comments, which express opinions, narratives, and reactions. To ensure the dataset's quality and modeling readiness, several preprocessing steps were undertaken. First, records with missing user information, empty comment fields, or duplicate entries were removed. The textual content then underwent rigorous normalization, including the removal of embedded URLs and non-alphabetic characters, conversion of all text to lowercase, elimination of standard English stopwords (using the NLTK stopword list), and lemmatization via the WordNet lemmatizer in NLTK to reduce words to their base forms. This process ensured syntactic consistency, reduced vocabulary sparsity, and prepared the text for machine learning applications. Following text normalization, each comment was annotated with an emotion label—positive, neutral, or negative—using a dual-method approach: a rule-based VADER classifier for fast preliminary sentiment scoring, and a transformer-based RoBERTa model (cardiffnlp/twitter-roberta-base-sentiment) for accurate, context-aware classification. The final structured dataset retained the following fields: *source_user, target_user, post_id, roberta_label, and text_processed.*

This sentiment-enriched interaction graph forms the empirical basis for downstream tasks, including graph neural network (GNN) classification, emotion diffusion analysis, and the evaluation of intervention strategies. A sample of the structured dataset is shown in Figure 3.

Figure 3:Real world dataset

| source_use | target_use | text | post_id | processed_ | vader_scor | vader_labe | roberta_la | roberta_score |
|---|---|---|---|---|---|---|---|---|
| WorldNews | socialist | > German manufacturer Rheinmetall is massively expanding its ammunition production.<br><br>> The production capacity of 155mm artillery shells will increase fifteenfold by 2027 compared to the level at | 1jwh3sq | german mar | 0.743 | positive | LABEL_2 | 0.5849699378013611 |

### 4.3 Real Reddit Emotion Diffusion Graph

To model real-world emotion diffusion, a directed graph based on Reddit user interactions using the NetworkX library is constructed:

Each node represents a unique Reddit user, and each directed edge denotes a reply or comment from one user to another (i.e., source_user → target_user).The edges are annotated with emotion labels (positive, neutral, negative) inferred from RoBERTa sentiment classification.

To explore the structural patterns of emotion propagation, we visualized two types of subgraphs:

- A Top-50 user graph, constructed from the most active users based on node degree;

- A thread-level diffusion graph, filtered by individual post_id, capturing the emotional flow within specific discussion topics.

These graph structures serve as the basis for downstream graph neural network modeling and emotion propagation analysis. The network is shown in figure 4:

Figure 4:Reddit emotion diffusion network(Top 50)

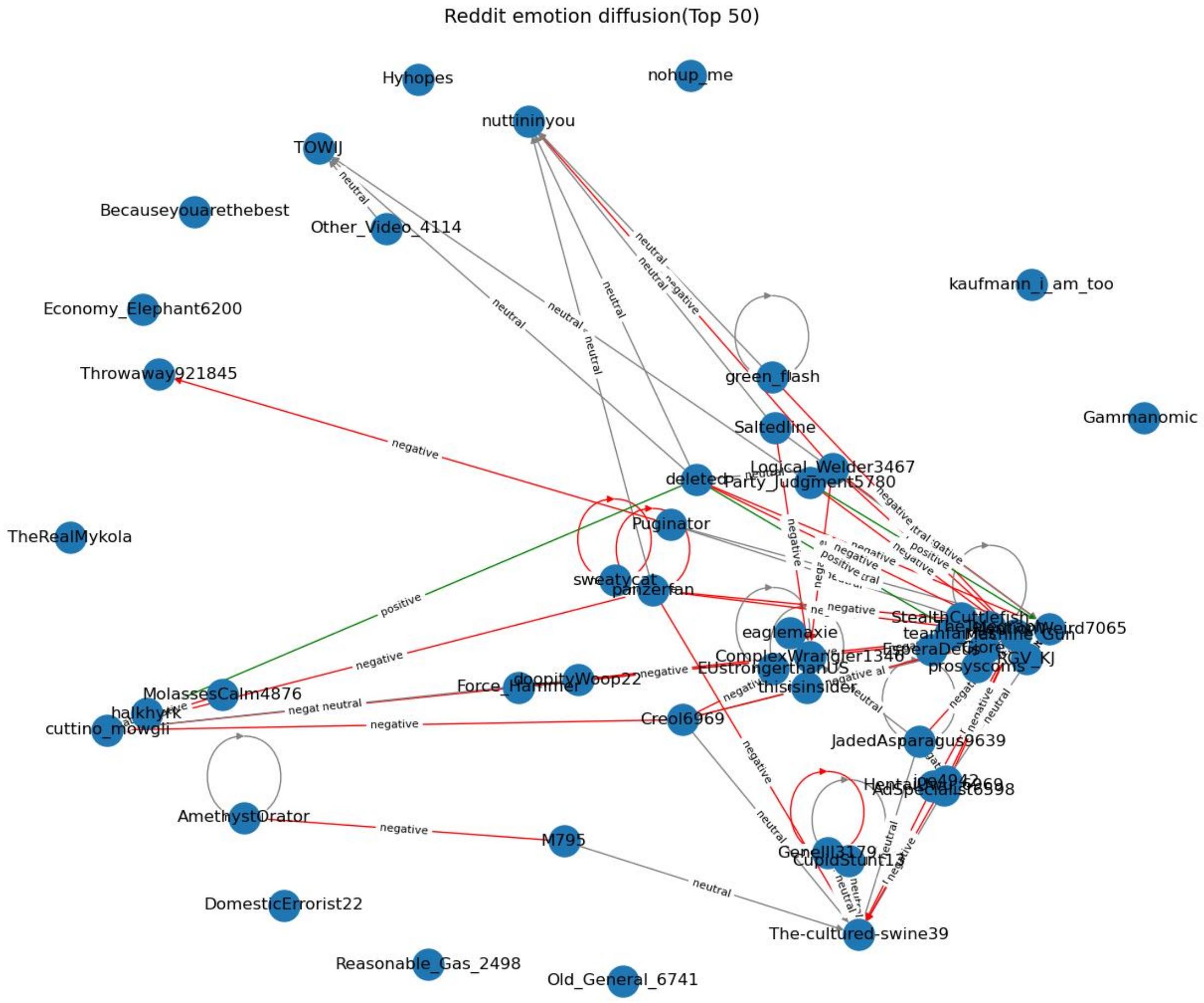

Two emotion diffusion networks are constructed: a real graph (*G_real*) extracted from Reddit political discussions and a simulated graph (*G_sim*) generated using the DeepSeek-Chat model. A comparison of their topological properties reveals stark structural contrasts.

The *G_real* graph comprises approximately 9,000 unique users and over 13,000

directed edges, demonstrating rich connectivity and structural complexity. Its degree distribution is long-tailed, indicating the presence of highly active users (hubs) who engage repeatedly across threads. The network exhibits frequent bidirectional exchanges, feedback loops, and a high clustering coefficient, all of which reflect the nature of dynamic, multi-turn discussions typical in real online forums. Notably, clear community structures emerge around contentious topics, highlighting the organic formation of opinion-based subgroups.

In contrast, the *G_sim* graph, derived from 100 DeepSeek-simulated reply chains (A→B→C→D), contains only 400 nodes and 300 edges. Its structure is a collection of isolated, linear chains, with no overlapping users across chains. Every user appears only once, and no cycles or re-engagements occur, resulting in a clustering coefficient of zero. All edges are unidirectional, and the network lacks the interconnectedness and emergent organization seen in the real graph. These findings highlight a key limitation of current LLM-based simulations: while capable of generating plausible individual responses, they fail to replicate the interactive dynamics and structural richness of real-world social platforms.

The sentiment propagation patterns also diverge significantly. In *G_real*, emotional transitions are frequent, with reply chains often exhibiting sentiment shifts—including escalation, disagreement, and affective divergence—across turns. This emotional heterogeneity is consistent with the nature of spontaneous political discourse, where responses are shaped by disagreement, sarcasm, or reframing.

By contrast, *G_sim* displays sentiment monotonicity. Replies tend to preserve the emotional tone of the initial prompt, with minimal variation across chain steps. This likely reflects the deterministic influence of emotion-conditioned prompting and the absence of interpersonal nuance in synthetic interactions.

Figure 5 (Real Reddit Graph) visually depicts a dense, highly entangled network, with overlapping chains and complex sentiment flows. In contrast, Figure 6 (LLM-Simulated Graph) shows a sparse and disconnected structure, composed of non-interacting linear chains, revealing the lack of conversational depth and emergent behavior in the simulation.

Figure 5: Real Reddit Emotion Diffusion Graph

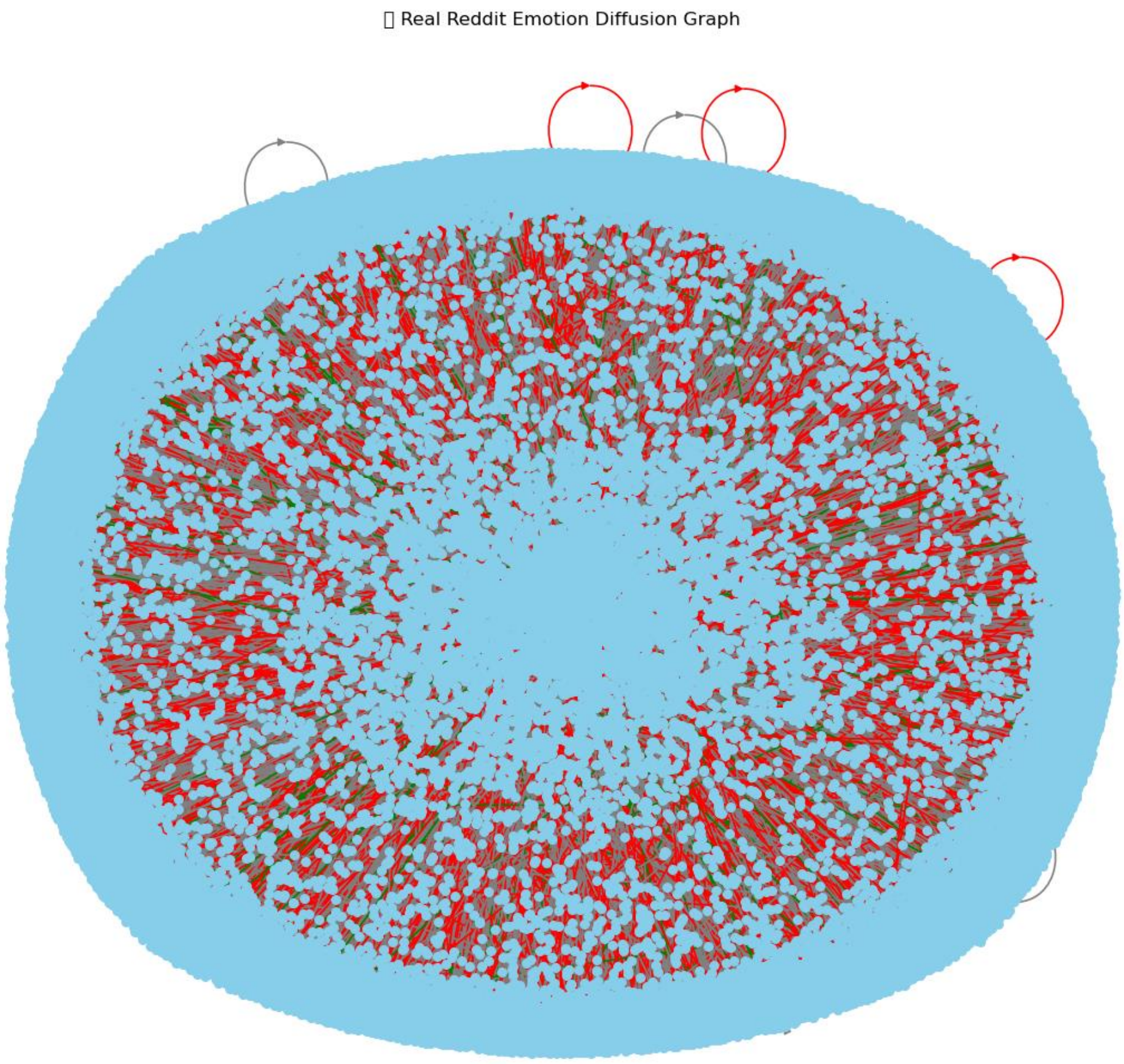

Figure 6:LLM-simulated Diffusion graph

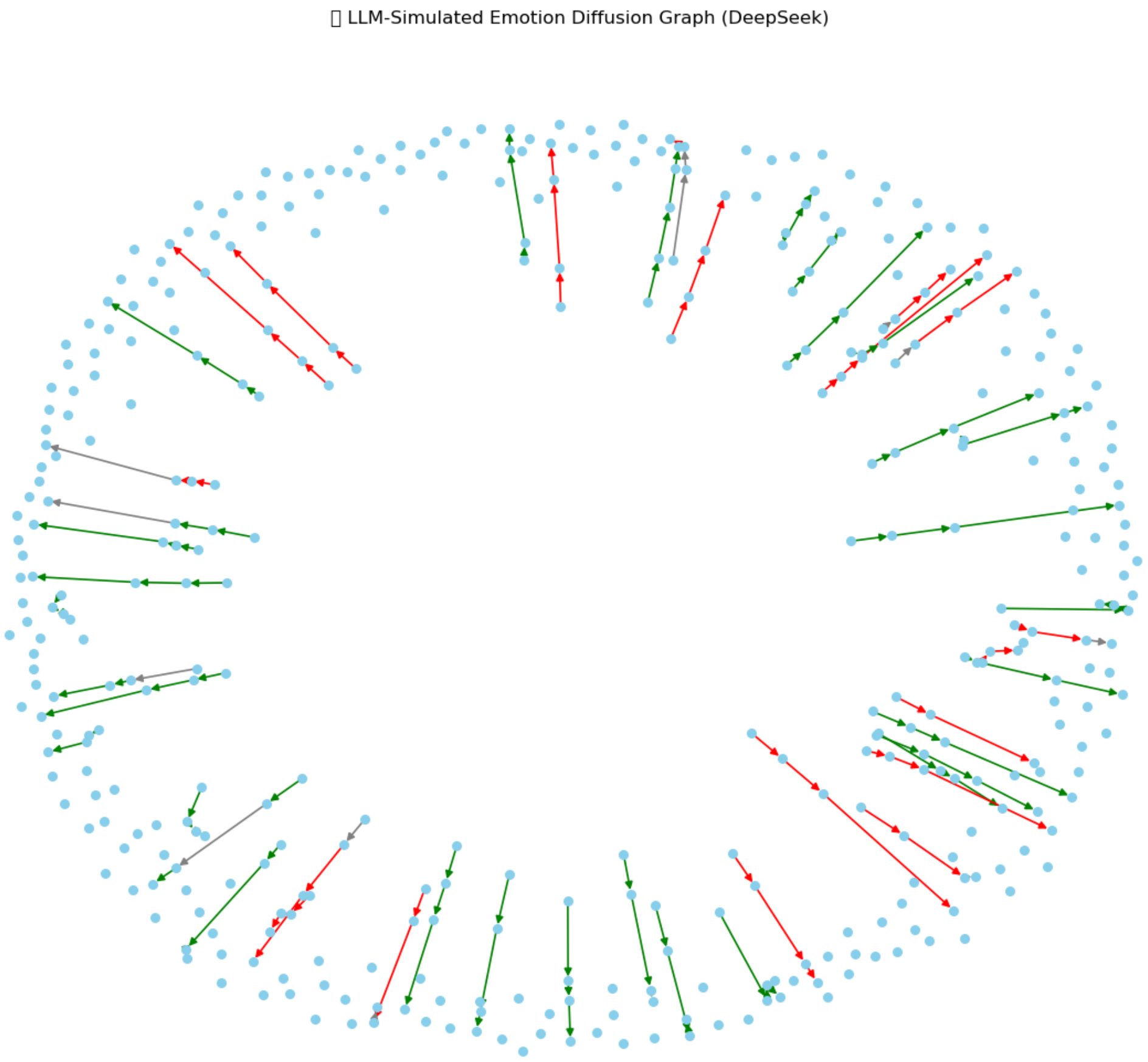

With an expanded simulated graph containing 400 sentiment-labeled edges, we retrained a GNN classifier to evaluate performance under increased emotional and structural diversity. The model achieved 75.5% accuracy and a macro F1 score of 0.621, showing robust performance across all sentiment classes. Notably, the F1 scores for both *negative* and *neutral* categories exceeded 0.55, suggesting the model effectively leveraged structural signals and sentiment context, even in constrained chain-based interactions. Compared to previous results, this version better approximates class balance and emotional variability.

## 5.Discussion

This study shows that large language models (LLMs) have the ability to simulate the process of social media emotion propagation to a certain extent. By generating subsequent replies based on the sentiment of the previous comment, LLMs are able to construct chains of interactions with emotional consistency, superficially reproducing the patterns of emotional responses in a community. This opens up the possibility of constructing controlled simulation environments in the absence of real data or limited by privacy constraints, and is particularly suitable for testbeds to study emotion diffusion mechanisms and intervention strategies.

However, there are also obvious limitations of our approach. First, the generated simulation graphs are oversimplified in terms of structure. Compared to the complex network structure in real communities such as Reddit, where feedback loops, cross-interactions, and aggregation of active users exist, the graphs generated by LLM only contain non-overlapping linear chains, which cannot truly reflect natural dialogue behaviors such as multi-party participation, counter-arguments, and group formation. This structural limitation also weakens the ability to reproduce emergent features in social networks.

Second, from the perspective of sentiment propagation, the simulation graph shows a single change in sentiment. Most replies remain emotionally consistent with the initial posting and lack the emotional turnaround, conflict, or escalation process that is common in real discussions. This phenomenon may stem from the fact that the current LLM generation method mainly relies on instantaneous prompts and lacks the support of historical context or user memory.

Third, from the GNN classification experiments, although the overall accuracy of the model is moderate, it still faces challenges in dealing with unbalanced categories and homogeneous structures. The model performs well on the dominant emotion categories, but its prediction ability on a few categories such as 'neutral' and 'negative' is obviously insufficient, which indicates that the current training data and graph structure limit the breadth and generalization of emotion recognition.

To address the above problems, the following aspects can be improved in the future: firstly, allowing users to reappear in multiple chains, so as to construct a network structure closer to the real discussion, and promote the formation of feedback loops and cross-paths. Second, user memory mechanisms or dialogue history contexts can be introduced to enhance the coherence and diversity of sentiment generation. Third, reinforcement learning or fine-tuning can be used to optimise the generation strategy and encourage the generation of responses with more structural complexity and emotional heterogeneity.

In summary, although the current LLM simulation results can initially reproduce the basic features of emotion diffusion in some aspects, we still need to continue exploring and optimizing the structural restoration, emotion modelling and interaction depth in order to truly simulate the real social system.